\documentclass{article}

    \PassOptionsToPackage{numbers, compress}{natbib}
\usepackage[preprint]{neurips_2025}

\usepackage[utf8]{inputenc} % allow utf-8 input
\usepackage[T1]{fontenc}    % use 8-bit T1 fonts
\usepackage{url}            % simple URL typesetting
\usepackage{booktabs}       % professional-quality tables
\usepackage{amsfonts}       % blackboard math symbols
\usepackage{nicefrac}       % compact symbols for 1/2, etc.
\usepackage{microtype}      % microtypography
\usepackage{xcolor}         % colors
\usepackage{color, colortbl}
\usepackage{multirow}
\usepackage{graphicx}
\usepackage{amsmath}
\usepackage{amssymb}
\usepackage{mathtools}
\usepackage{amsthm}
\usepackage{mdframed}
\usepackage{relsize}
\usepackage{xspace}
\usepackage{wrapfig}
\usepackage{soul}
\usepackage{listings}
\usepackage{enumitem}
\setlist[itemize]{topsep=0pt}
\setlist[enumerate]{noitemsep, topsep=0pt}
\usepackage{xcolor,colortbl}

\definecolor{paleaqua}{rgb}{0.74, 0.83, 0.9}

\newcommand{\hlgreen}[1]{\sethlcolor{green!30}\hl{#1}}

\newcommand{\hlred}[1]{\sethlcolor{red!15}\hl{#1}}

\definecolor{dark_green}{rgb}{0.0, 0.5, 0.0}
\definecolor{maroon}{cmyk}{0, 0.87, 0.68, 0.32}
\definecolor{halfgray}{gray}{0.55}
\definecolor{ipython_frame}{RGB}{207, 207, 207}
\definecolor{ipython_bg}{RGB}{247, 247, 247}
\definecolor{ipython_red}{RGB}{186, 33, 33}
\definecolor{ipython_green}{RGB}{0, 128, 0}
\definecolor{ipython_blue}{RGB}{64, 128, 128}
\definecolor{ipython_purple}{RGB}{170, 34, 255}
\newcommand{\freshstack}{FreshStack\xspace}

\definecolor{citecolor}{RGB}{34,139,34}
\usepackage[pagebackref=true,breaklinks=true,colorlinks,
citecolor=citecolor,bookmarks=false]{hyperref}

\title{\freshstack: Building Realistic Benchmarks for Evaluating Retrieval on Technical Documents}

\author{%
 Nandan Thakur$^{\ 1}$\thanks{Work done during Nandan's internship at Databricks.}
  \ \ \ \ \ \ \ \
  Jimmy Lin$^{\ 1}$
  \ \ \ \ \ \ \ \ \ \
  Sam Havens$^{\ 2}$ \ \ \ \\
  \ \ \ \ \ \ \ \textbf{Michael Carbin}$^{\ 2}$
 \ \ \ \ \ \ \
  \textbf{Omar Khattab}$^{\ 2}$ 
\ \ \ \ \ \ \ \ \ \ \ 
  \textbf{Andrew Drozdov}$^{\ 2}$ \ \ \ \ \ \ \ \ \\
{$^1$}{University of Waterloo, Canada} \ \ \ \ \ \ {$^2$}{Databricks, USA}
\\
\url{https://fresh-stack.github.io}
}

\begin{document}

\maketitle

\begin{abstract}
We introduce \freshstack, a holistic framework for automatically building information retrieval (IR) evaluation benchmarks by incorporating challenging questions and answers.
\freshstack conducts the following steps:
(1) automatic corpus collection from code and technical documentation,
(2) nugget generation from community-asked questions and answers, and
(3) nugget-level support, retrieving documents using a fusion of retrieval techniques and hybrid architectures.
We use \freshstack to build five datasets on fast-growing, recent, and niche topics to ensure the tasks are sufficiently challenging. On \freshstack, existing retrieval models, when applied out-of-the-box, significantly underperform oracle approaches on all five topics, denoting plenty of headroom to improve IR quality. 
In addition, we identify cases where rerankers do not improve first-stage retrieval accuracy (two out of five topics) and oracle context helps an LLM generator generate a high-quality RAG answer.
We hope \freshstack will facilitate future work toward constructing realistic, scalable, and uncontaminated IR and RAG evaluation benchmarks. 
% \freshstack datasets are available at: \href{https://fresh-stack.github.io}{https://fresh-stack.github.io}.
\end{abstract}

\section{Introduction}
\label{introduction}

% Talk briefly about RAG.
Retrieval-augmented generation (RAG) is a popular technique to enhance traditional information retrieval (IR) capabilities with language model generation. 
RAG systems use large language models (LLMs) to generate long-form responses~\cite{guu:2020, lewis:2020, izacard:2021, borgeaud:2022}, grounded in the information available from retrieved documents \cite{khandelwal:2020, lewis:2020, gao:2023, liu:2024}.
Despite its wide usage, evaluating RAG remains incredibly challenging. 
Existing IR and RAG benchmarks are not well-suited for evaluation, as these are outdated and highly limited. 
In particular, we observe three major issues in existing benchmarks:

\begin{itemize}[leftmargin=0.3cm]
\item \textbf{Lack of realistic, open-ended questions}: Existing datasets contain purely extractive short answers (e.g., Natural Questions \cite{kwiatkowski:2019}, TriviaQA \cite{joshi:2017}) or crowd-sourced questions (e.g., HotPotQA \cite{yang:2018}). 
A limited number of datasets capture ``natural'' human-asked questions, i.e., MS MARCO \cite{nguyen:2016} or Natural Questions \cite{kwiatkowski:2019}, but unfortunately, brief and straightforward questions are inserted into a search box, failing to represent the complex questions that real users might pose to RAG systems.

\item \textbf{Artificially easy}: RAG represents an \emph{approach} rather than a \emph{problem}. 
Real users require systems capable of grounded question answering, i.e., responding to specialized questions by referencing knowledge from a document corpus. 
Consequently, datasets constructed by design to be solvable via retrieval often fail to encode challenges faced in RAG applications.

\item \textbf{Static and unspecialized}: After sourcing questions and answers, a benchmark becomes at the risk of (1) \emph{contamination}, if current LLMs are trained on the same set of documents or questions, (2) \emph{overfitting}, when systems inevitably saturate by repeated internal or external leaderboarding (e.g., BEIR \cite{thakur:2021}), and (3) \emph{staleness}, when questions or answers are not refreshed and become outdated. 
\end{itemize}

\begin{figure*}[t!]
        \centering
        \resizebox{\textwidth}{!}{
        \includegraphics[width=1.0\textwidth,trim={25 20 10 20}]{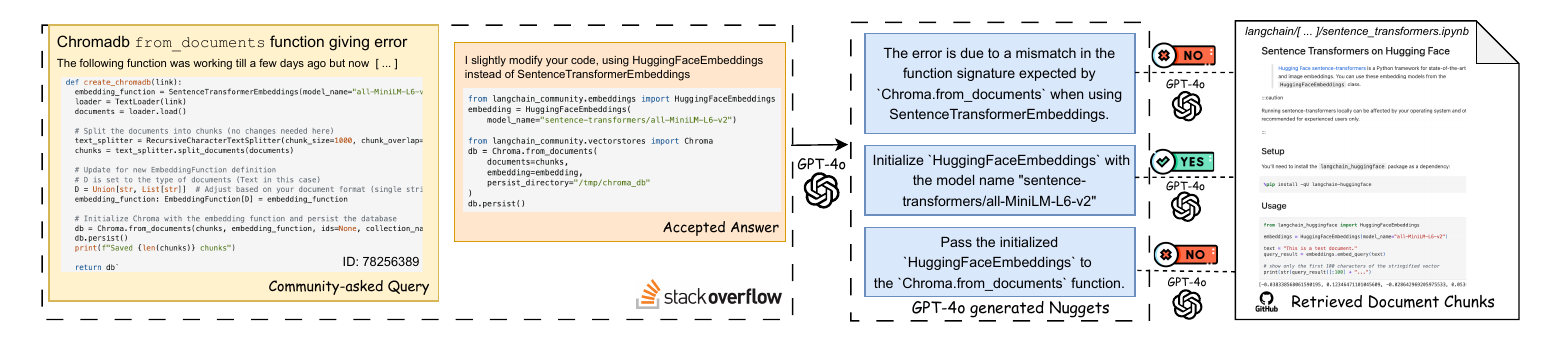}
        }
    \caption{A data instance from LangChain generated with \freshstack. The question and answer pair is sourced from Stack Overflow. The pair is provided to GPT-4o to generate nuggets, highlighting necessary facts in the answer. Next, code snippets and technical documents from multiple GitHub repositories (e.g., Jupyter Notebook) are chunked, processed, and pooled for each question. Finally, each pooled document chunk is judged with GPT-4o for binary relevance (either yes or no) at a nugget-level, i.e., whether the document factually supports the information present in each nugget.}
    \label{fig:langchain}
\end{figure*}

\noindent A realistic benchmark should measure model generalization on niche domains, and continue to update. Additionally, it must capture the complexity of human-generated queries---such as multi-hop reasoning~\cite{yang:2018}, code understanding~\cite{jimenez:2024}, or specialized terminology---rather than relying on artificially easy questions. This drives the robustness of systems in answering questions in evolving public libraries or private code bases \cite{zhang-etal-2023-repocoder, jimenez:2024}, a company's internal forum~\cite{raina-gales-2024-question}, or technical troubleshooting~\cite{soman:2024, pu:2025, chen:2025}.

In our work, we introduce \textbf{\freshstack}, a holistic framework for constructing realistic datasets on niche and challenging domains, seeking to avoid contamination due to (perpetual) recency. 
Using \freshstack, we construct an evaluation benchmark on five niche topics sourced from community-asked questions and answers on Stack Overflow and a corpus containing code snippets and technical documents from public GitHub repositories. 
The framework contains three major steps:~
(1) Automatic corpus collection (Section \ref{sec:corpus-collection}) with technical documents chunked and sourced from several GitHub repositories.
(2) Nugget generation (Section \ref{sec:nuggetization}) with GPT-4o using community-asked questions and answers in Stack Overflow.
(3) Nugget-level support (Section \ref{sec:nugget-support}) with GPT-4o on document chunks, retrieved from a fusion of retrieval techniques and hybrid architectures.

We investigate three research questions in our work to provide insights on \freshstack: (RQ1) How to construct challenging evaluation datasets from real user-asked questions? (RQ2) How do LLMs act as an assessor for nugget generation on community-asked questions \& answers and nugget-level support with retrieved documents? (RQ3) How do state-of-the-art retrieval models, rerankers, and LLM generators perform on IR and RAG evaluation benchmarks generated with \freshstack?

We calibrate the automatic stage in \freshstack with GPT-4o using a machine learning (ML) expert, assessing the quality of nugget generation and nugget-level support for one of the topics (LangChain). 
Our results show that GPT-4o-generated nuggets capture crucial information required to answer the question, and GPT-4o precisely labels support at a nugget level. For pooling, we compare oracle (having access to the answer) and inference (relying only on the question) settings, finding that question decomposition and nugget generation outperform other techniques, respectively.

Beyond pool construction, we evaluate retrieval and rerankers in the document retrieval setting using only the Stack Overflow question. 
Retrieval models drastically underperform oracle systems on all five topics, showing a high headroom for improvement. In addition, ensemble fusion outperforms individual models, indicating that diversity in models enhances retrieval, and rerankers provide clear benefits in some but not all topics. Finally, we evaluate answer generation, where the oracle context assists the LLM generator to provide a high-quality RAG answer. \freshstack is a general framework and can be applied to any domain of a similar structure. 
Overall, we hope the framework serves as a testbed for future work to develop challenging benchmarks for evaluating RAG systems.

\section{Related Work}

\textbf{Retrieval-augmented generation.} RAG has been widely used to avoid ``hallucinations'' \cite{zhang:2023} seen with LLMs when handling knowledge-intensive tasks \cite{kandpal:2023}. 
RAG reduces factually incorrect generated content, leading to adoption in various commercial systems, e.g., Bing Search or Google AI Overviews. 
Existing IR and RAG benchmarks are stale, evaluating on academic question answering datasets \cite{gao:2023, rosenthal:2024, ram:2023, su2025bright, xiao2024rarbreasoningretrievalbenchmark}, or are not challenging, being constructed for RAG~\cite{yang:2024, chen:2024, li:2024b, niu:2024, su2025bright, krishna-etal-2025-fact}. 
A limited number of datasets refresh over time to avoid LLM decontamination~\cite{kasai:2023, vu:2024, shao:2024, white2025livebench}, however, these contain easy and unrealistic questions. In contrast, \freshstack generates niche and challenging datasets, which can refresh over time and are not constructed specifically for RAG.

\textbf{Code-based benchmarks.}~Neural code generation \cite{lu2021codexglue} requires LLMs to generate code from scratch for generic programming questions. 
One popular benchmark is SWE-Bench~\cite{jimenez:2024}, which evaluates whether LLMs can generate code changes for GitHub pull requests (PRs) in popular public repositories.
Similarly, CodeSearchNet \cite{husain:2019}, COIR \cite{li:2024}, LiveCodeBench~\cite{jain:2024}, and CodeRAG-Bench \cite{wang:2024} focus on the evaluation of high-level programming problems on popular public repositories. 
In contrast, in \freshstack, we focus on assisting developers, from a novice to a domain expert, by providing real-time answers on fast-growing and recent topics such as LangChain (introduced in 2023) by referencing technical documentation in GitHub repositories.

\textbf{Stack Overflow datasets.} \freshstack is \emph{not} the first dataset to use Stack Overflow for retrieval. However, the evaluation setting of retrieving canonical documents from GitHub repositories remains under-explored.
Existing datasets such as CQADupstack \cite{hoogeveen:2015}, LoTTE \cite{santhanam:2022}, and Stack Overflow-QA \cite{li:2024} follow a different task, to retrieve the relevant answer snippet given the question asked by a real user on Stack Overflow. 
The closest setting similar to \freshstack is found in Neural Code Search~\cite{li:2019b}, which incorporates public documentation and code snippet examples from GitHub as the corpus to answer questions asked by real users on popular programming topics such as Android. 

\begin{wrapfigure}{r}{0.5\textwidth}
    \resizebox{0.5\textwidth}{!}{
    \includegraphics[width=\textwidth,trim={25 20 10 20}]{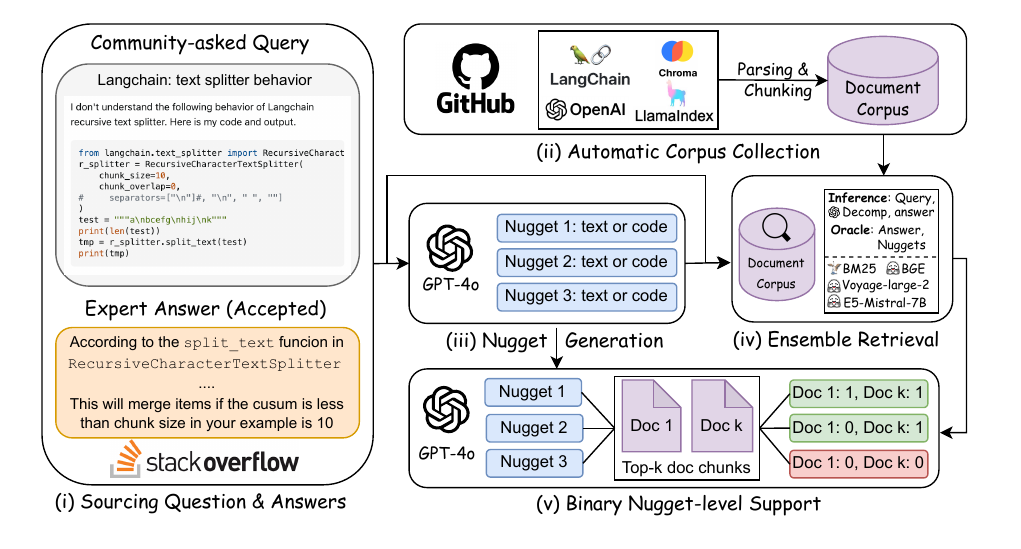}
    }
    \caption{The \freshstack framework: (1) Stack Overflow questions and answers are sourced for recent and niche topics. (2) GitHub repository documents are collected and chunked to form the corpus (for each topic). (3) Nuggets or atomic facts within the question and answer are generated with GPT-4o. (4) Ensemble techniques and models retrieve documents, which construct our document judgment pools. (5) GPT-4o evaluates support for every document-nugget pair as a binary judgment. \vspace{-15mm}}
    \label{fig:overview}
\end{wrapfigure}

\section{The \freshstack Framework}

The framework involves five major stages to construct an evaluation dataset (as highlighted in \autoref{fig:overview}). \freshstack includes three major design choices:
\begin{enumerate}[leftmargin=0.5cm]
    \item A general framework that can be extended to different domains without manual effort.
    \item Adding recent and niche topics actively discussed in computer programmer communities such as Stack Overflow.
    \item Sourcing community-asked questions and answers, to make our evaluation challenging, requiring domain expert knowledge to answer them correctly.
\end{enumerate}

\textbf{Stack Overflow}~is an online question answering platform for computer programmers.  
Users ask questions about a particular topic and provide a description (often a code snippet with the error message) with Stack Overflow tags.  
Questions and answers are also tagged by topics, allowing for easy retrieval of topic-wise questions.

\subsection{Stack Overflow Topic and Question Selection}
For topic selection, we target niche and recent topics introduced on Stack Overflow from 2023 onward (frequency in \autoref{fig:freshstack-questions}), containing a minimum of 50 posts. 
We sort all topics using the overall number of posts and curate five topics starting from the highest (LangChain) to the lowest frequency (Yolo v7~\&~v8), covering different domains and sufficiently different from each other. 
Each topic represents a unique domain, e.g., Computer Vision (CV) or Machine Learning (ML).

\textbf{Questions \& Answers.} We extract relevant posts and answers from the Stack Overflow XML data archive (dated October 2024).\footnote{The Stack Overflow XML data archive (CC BY-SA license) is updated once every quarter: \url{https://meta.stackexchange.com/questions/401324/announcing-a-change-to-the-data-dump-process}} 
We scan the archive to pick all the relevant posts as questions containing the required tag and further filter answer posts to the questions. 
Finally, we keep questions with an \emph{accepted answer}, prioritizing precision over quantity in retaining questions with only high-quality answers accepted and upvoted in the Stack Overflow community.

\subsection{Automatic Corpus Collection with GitHub Technical Documents}\label{sec:corpus-collection}
Answering a higher percentage of questions requires a robust set of corpora from \emph{multiple sources}. 
For instance, addressing issues in LangChain may require ChromaDB GitHub documentation to resolve errors related to its usage. 
In our work, we build a different document corpus per topic by combining multiple repositories as sources (we list each repository per topic in \autoref{tab:github_repositories}). 

\textbf{Stack Overflow Tags.} We analyze each tag frequency from Stack Overflow to select a relevant GitHub repository to be included in the document corpus.
This involves identifying top-k co-occurring tags, where k is the threshold balancing question coverage with indexing costs. 
Some tags are generic, such as \texttt{Python}, whereas others are specific, such as \texttt{LangChainJS}. 
Filtering the tags to keep only a subset of repositories does not degrade the dataset quality. 
We manually verify each GitHub repository for each tag, with plans to automate this procedure in the future.

\textbf{Chunking \& Indexing.} We clone the latest branch of the GitHub repository in a local workspace and parse files as a tree structure. 
Each file (either a text document or code snippet) is chunked into small sections containing a maximum of \textbf{2048 tokens}, skipping non-text formats.\footnote{We skip indexing images, videos, .bin, .csv, and audio files or unrecognized file formats.}~The GitHub filepath serves as the document identifier, with additional chunk details encoded in the identifier. Finally, we combine all chunks into a single corpus, prefixing all document identifiers with the repository name to identify the common files in each repository separately (e.g., LICENSE or requirements.txt).

\subsection{Nuggetization or Nugget Generation}\label{sec:nuggetization}

A nugget is a core concept or an atomic fact essential in a system's response. The term nugget was informally referred to as SCU (summary content units) as clauses appearing in model summarization \cite{nenkova:2004} and later formalized as ``information nugget'' for evaluating long-form answers \cite{voorhees:2003, lin:2005, lin:2006, pavlu:2012}. \emph{Nuggetization} refers to constructing or generating nuggets from information-dense text. 
The procedure decomposes a verbose answer into key atomic facts or essential components, aiding evaluation. 
More recently, with the onset of RAG, nugget-based evaluation has renewed interest with LLMs for factual accuracy assessment in long answers \cite{arabzadeh:2024, mayfield:2024, farzi:2024, pradeep:2024, pradeep:2024a, pradeep2025greatnuggetrecallautomating}.

\textbf{Nuggetization.} We automatically generate nuggets from Stack Overflow question-answer pairs using GPT-4o \cite{openai_gpt4o}, avoiding the cumbersome procedure of manual nugget construction~\cite{pradeep:2024}. 
LLM-based nugget generation has been explored in the TREC 2024 RAG track\footnote{TREC 2024 RAG track: \url{https://trec-rag.github.io/}} \cite{pradeep:2024, pradeep:2024a, pradeep2025greatnuggetrecallautomating} and in multiple works \cite{dietz:2024, farzi:2024}. 
Separately, we experimented with prompting techniques and found that grading notes style prompts~\cite{grading-notes:2024} provided parseable and high-quality nuggets in our experiments. An example of GPT-4o-generated nuggets for a question in LangChain is shown in \autoref{tab:example}.

\subsection{Retrieval: Oracle \& Inference Setting}\label{sec:retrieval}
A RAG evaluation dataset requires questions, answers, and a corpus with documents, which helps support facts in the answer. 
In this stage, we retrieve a list of highly relevant (unjudged) documents from the document corpus and construct judgment pools. 
Since, we are constructing an \emph{evaluation dataset} and we have \emph{curated answers} for questions, we retrieve documents using two methods: (1) \textbf{Inference}, relying only on the question and automatic approaches, and (2) \textbf{Oracle}, relying on the gold answer or list of nuggets, to pool diverse documents for relevance judgment in the next stage.

\textbf{Retrieval Settings.} We experiment with multiple systems to increase diversity in our judgment pools. First, we experiment with two techniques in the \emph{inference setting}:\ (i) GPT-4o Sub-Questions: we decompose the original question and generate synthetic sub-questions with GPT-4o, similar to \citet{rosset:2024}, concatenated together to retrieve documents, and (ii) GPT-4o Closed Book Answer: we generate a closed-book answer for the original question with GPT-4o, similar to HyDE \cite{gao:2023c}, and use the closed-book answer to retrieve documents.
Next, in the \emph{oracle setting}, we experiment with:\ (i) Stack Overflow Answer: we use the curated Stack Overflow answer as the question to retrieve documents, and (ii) Stack Overflow Nuggets: we use the list of GPT-4o-generated nuggets (Section \ref{sec:nuggetization}), concatenated as the question to retrieve documents.

\textbf{Retrieval Models.} We experiment with five different code and text-aware retrieval models: (i) \textbf{BM25}, a strong lexical baseline in BEIR \cite{thakur:2021}. We utilize the default implementation available in Pyserini \cite{lin:2021}.
(ii) \textbf{BGE (Gemma-2)} \cite{chen:2024} a dense retriever model\footnote{BGE Gemma-2: \url{https://huggingface.co/BAAI/bge-multilingual-gemma2}} fine-tuned on the backbone architecture of Gemma-2 (9B) \cite{gemma2:2024} with an embedding size of 3584 and 8K context length. 
(iii) \textbf{E5 Mistral (7B)} \cite{wang:2024b} is a dense retriever model\footnote{E5 Mistral 7B: \url{https://huggingface.co/intfloat/e5-mistral-7b-instruct}} based fine-tuned on the backbone of Mistral 7B \cite{jiang:2023} with 32 layers and embedding size of 4096.
(iv) \textbf{Voyage-large-2}\footnote{Voyage-large-2: \url{https://docs.voyageai.com/docs/embeddings}} is a proprietary and general-purpose embedding model optimized for retrieval quality, with a context length of 16K tokens and embedding size of 1536.
(v) \textbf{Fusion}, a hybrid retrieval strategy combining the four individual models, normalizing and summing up the top 100 documents and their scores from each model.

\subsection{Nugget-Level Support Assessment with Retrieved Documents}\label{sec:nugget-support}
Traditionally, relevance judgments are conducted on selected pools of retrieved documents, i.e., where a human assessor judges the relevance of the question with each provided document. 
Due to computational costs, recent studies experiment with an LLM judge (instead of a human assessor) for relevance judgments in IR~\cite{faggioli:2023, thomas:2024, upadhyay:2024a, upadhyay:2024b, rahmani:2024}. 
Questions in existing IR datasets are traditionally short, making it easier to judge document relevance. In contrast, questions in the \freshstack dataset are long and elaborate (between 350--500 tokens in length), containing a mixture of text, code snippets, or outputs, making it challenging to judge question-document relevance directly \cite{damessie:2016}.
For instance, a document may answer a major problem presented in the question, address only part of the question, or contain relevant references and examples, and we need to translate this into a relevance score.

\begin{table*}[t!]
\centering
\resizebox{\textwidth}{!}{
\begin{tabular}{l|l|rrcc|cc|cc|cc}
\toprule
\multicolumn{1}{l}{\multirow{2}{*}{\textbf{Topic}}} &
\multicolumn{1}{l}{\multirow{2}{*}{\textbf{Domain}}} &
\multicolumn{4}{c}{\textbf{Dataset Count}} & \multicolumn{2}{c}{\textbf{Avg. Length}} & \multicolumn{2}{c}{\textbf{\%~Containing Code}} & \multicolumn{2}{c}{ \textbf{Relevance Judgments}} \\ 
\cmidrule(lr){3-6} \cmidrule(lr){7-8} \cmidrule(lr){9-10} \cmidrule(lr){11-12}

\multicolumn{1}{l}{} &
\multicolumn{1}{l}{} &
\textbf{\#Q} & \textbf{\#Docs} & \multicolumn{1}{c}{\textbf{\#GH}} &
\multicolumn{1}{c}{\textbf{Avg. N/Q}} & 
\textbf{Query} & \multicolumn{1}{c}{\textbf{Answer}} & 
\textbf{Query} & \multicolumn{1}{c}{\textbf{Answer}} &
\textbf{Rel. Docs/N} & \textbf{Rel. Docs/Q} \\ \midrule
\textbf{LangChain} & Machine Learning (ML) & 203 & 49,514 & 10 & 3.1 & 473.4 & 233.8 & 83.3\% & 62.1\% & 5.7 & 10.9 \\
\textbf{Yolo v7 \& v8} & Computer Vision (CV) & 57 & 27,207 & 5 & 3.5 & 497.1 & 191.7 & 70.2\% & 71.9\% & 3.9 & 7.4 \\
\textbf{Laravel 10 \& 11} & Backend Development & 184 & 52,351 & 9 & 3.0 & 474.4 & 155.5 & 43.5\% &  51.1\% & 3.2 & 6.0 \\
\textbf{Angular 16, 17 \& 18} & Front-end Development & 129 & 117,288 & 4 & 3.2 & 463.3 & 215.1 & 69.8\% & 57.4\% & 4.4 & 8.7 \\
\textbf{Godot4} & Game Development & 99 & 25,482 & 6 & 3.3 & 350.4 & 263.0 & 52.5\% & 52.5\% & 2.9 & 5.9 \\ \bottomrule
 
\end{tabular}
}
\caption{\freshstack dataset statistics; Dataset count measures the number of queries, documents, GitHub repositories, and average nuggets per query; Avg.~length measures the average text lengths (length distribution in \autoref{fig:token-distribution}); \% containing code measures the fraction of queries and answers with code snippets; Relevance judgments measure relevant documents per nugget and per query.}
\label{tab:dataset-statistics}
\end{table*}

\textbf{Nugget-level Support.} Instead of relying on traditional relevance assessments, we simplify the judgment procedure for GPT-4o and evaluate whether a document supports information (or contains) provided by a nugget. 
A reminder that a nugget highlights an essential fact of the Stack Overflow question or answer. 
Judging document relevance at a nugget level is effective as nuggets are factual and short information snippets, reducing the ambiguity often seen during traditional relevance judgments. 
To reduce computational costs, we evaluate top-k documents (a maximum $k=20$) together with the list of all nuggets for a question in a single inference call ($n + k$). We evaluate support judgment with GPT-4o using a chain-of-thought prompt~\cite{wei:2023}.

\section{Dataset Statistics \& Evaluation}

Completing previous stages, we employ two additional post-processing steps to ensure high-quality question and answer pairs remain in the dataset, sacrificing the overall dataset size. 
In the first step, we remove unsupported questions, i.e., questions that do not contain even a single relevant document; this removes, on average, \textbf{11.8\%} of the total questions.\footnote{Future work may include these questions as they are potentially valuable to answer, and better retrieval systems may be able to find relevant documents.} 
In the next step, we aggressively filter by removing questions containing at least one unsupported nugget, i.e., a nugget not supported by any documents, reducing on average \textbf{34.2\%} of the total questions.

\subsection{Dataset Statistics}
\freshstack datasets covers five domains for programmers: machine learning, computer vision, backend, front-end, and game development, all listed in \autoref{tab:dataset-statistics}.
Stack Overflow topics such as LangChain were introduced in 2023, whereas others, like Laravel or Angular, have questions about the latest versions (e.g., Laravel 10 \& 11). 
Each topic has at least 50 questions, all asked between January 2023 and June 2024 (timeline versus frequency shown in \autoref{fig:freshstack-questions}). The corpus has at least 25K documents sourced from 4--10 GitHub repositories (repositories listed in \autoref{tab:github_repositories}). 
The questions are even longer than answers (distribution shown in \autoref{fig:token-distribution}), containing 350--500 tokens (computed using GPT-4o tokenizer), and at least 50\% of the questions and answers contain code snippets.
GPT-4o generates around 3--4 nuggets for each question. Each nugget supports at least 3 relevant documents, resulting in 5--6 relevant documents per question, for all topics.

\begin{table*}[t!]

\resizebox{\textwidth}{!}{
\begin{tabular}{l|l|ccc|ccc|ccc|ccc|ccc}
\toprule
\multicolumn{1}{l}{\multirow{2}{*}{\textbf{Method}}} & \multicolumn{1}{c}{\multirow{2}{*}{\textbf{Model}}} & \multicolumn{3}{c}{\textbf{LangChain}} & \multicolumn{3}{c}{\textbf{Yolo v7 \& v8}} & \multicolumn{3}{c}{\textbf{Laravel 10 \& 11}} & \multicolumn{3}{c}{\textbf{Angular 16, 17 \& 18}} & \multicolumn{3}{c}{\textbf{Godot4}} \\ 
\cmidrule(lr){3-5} \cmidrule(lr){6-8} \cmidrule(lr){9-11} \cmidrule(lr){12-14} \cmidrule(lr){15-17}
\multicolumn{1}{l}{} &
\multicolumn{1}{l}{} & $\alpha$N@10 & C@20 & \multicolumn{1}{c}{R@50} & $\alpha$N@10 & C@20 & \multicolumn{1}{c}{R@50} & $\alpha$N@10 & C@20 & \multicolumn{1}{c}{R@50} & $\alpha$N@10 & C@20 & \multicolumn{1}{c}{R@50} & $\alpha$N@10 & C@20 & R@50 \\ \midrule
\rowcolor{paleaqua} \multicolumn{17}{l}{\textit{\textbf{Inference Setting}: Using a variant of the Stack Overflow question for retrieval of documents within the corpus}} \\ \midrule

% gpt-4o sub-questions row
 \multirow{5}{*}{\shortstack[l]{GPT-4o \\Sub -\\Questions}} 
 & BM25 & 0.228 & 0.495 & 0.249 & 0.150 & 0.427 & 0.328 & 0.349 & 0.656 & 0.464 & 0.307 & 0.666 & 0.378 & 0.154 & 0.326 & 0.211 \\
 & BGE (Gemma-2) & 0.220 & 0.561 & 0.324 & 0.220 & 0.554 & 0.367 & 0.407 & 0.727 & 0.585 & 0.360 & 0.707 & 0.459 & 0.240 & 0.532 & 0.382 \\
 & E5 Mistral (7B) & 0.262 & 0.613 & 0.362 & 0.266 & 0.593 & 0.484 & 0.306 & 0.643 & 0.528 & 0.305 & 0.617 & 0.397 & 0.220 & 0.461 & 0.349 \\
 & Voyage-large-2 & 0.270 & 0.563 & 0.329 & 0.213 & 0.526 & 0.370 & 0.366 & 0.687 & 0.552 & 0.344 & 0.69 & 0.449 & 0.260 & 0.594 & 0.473 \\
 & Fusion (4 models) & \textbf{0.322} & \textbf{0.708} & \textbf{0.475} & 0.305 & 0.665 & 0.489 & \textbf{0.478} & \textbf{0.763} & \textbf{0.662} & \textbf{0.428} & \textbf{0.817} & \textbf{0.584} & \textbf{0.290} & \textbf{0.598} & \textbf{0.526} \\ \midrule

% gpt-4o answer row
\multirow{5}{*}{\shortstack[l]{GPT-4o \\Closed \\Book\\Answer}}
 & BM25 & 0.256 & 0.520 & 0.273 & 0.286 & 0.554 & 0.431 & 0.376 & 0.655 & 0.495 & 0.293 & 0.542 & 0.332 & 0.241 & 0.473 & 0.349 \\
 & BGE (Gemma-2) & 0.181 & 0.467 & 0.263 & 0.271 & 0.599 & 0.473 & 0.360 & 0.694 & 0.539 & 0.242 & 0.525 & 0.338 & 0.187 & 0.454 & 0.358 \\
 & E5 Mistral (7B) & 0.198 & 0.471 & 0.277 & 0.239 & 0.511 & 0.364 & 0.188 & 0.458 & 0.384 & 0.179 & 0.430 & 0.267 & 0.151 & 0.318 & 0.237 \\
 & Voyage-large-2 & 0.220 & 0.500 & 0.301 & 0.247 & 0.557 & 0.495 & 0.317 & 0.658 & 0.524 & 0.227 & 0.461 & 0.338 & 0.253 & 0.510 & 0.454 \\
 & Fusion (4 models) & 0.275 & 0.630 & 0.432 & \textbf{0.356} & \textbf{0.686} & \textbf{0.578} & 0.420 & 0.738 & 0.641 & 0.290 & 0.582 & 0.470 & 0.288 & 0.538 & 0.492 \\ \midrule

\rowcolor{paleaqua} \multicolumn{17}{l}{\textit{\textbf{Oracle Setting}: Using the Stack Overflow answer directly or its variants for retrieval of documents within the corpus}} \\ \midrule

% stack-overflow answer row
\multirow{5}{*}{\shortstack[l]{Stack \\Overflow \\Answer}}
 & BM25 & 0.461 & 0.726 & 0.428 & 0.481 & 0.756 & 0.574 & 0.511 & 0.774 & 0.588 & 0.469 & 0.751 & 0.521 & 0.325 & 0.565 & 0.397 \\
 & BGE (Gemma-2) & 0.290 & 0.625 & 0.367 & 0.390 & 0.815 & 0.604 & 0.472 & 0.814 & 0.675 & 0.346 & 0.690 & 0.481 & 0.341 & 0.718 & 0.561 \\
 & E5 Mistral (7B) & 0.331 & 0.671 & 0.430 & 0.315 & 0.683 & 0.509 & 0.260 & 0.634 & 0.488 & 0.291 & 0.570 & 0.412 & 0.277 & 0.546 & 0.434 \\
 & Voyage-large-2 & 0.385 & 0.700 & 0.432 & 0.405 & 0.703 & 0.589 & 0.439 & 0.791 & 0.641 & 0.371 & 0.680 & 0.477 & 0.371 & 0.626 & 0.541 \\
 & Fusion (4 models) & 0.484 & 0.821 & 0.619 & 0.546 & 0.854 & 0.788 & 0.564 & \textbf{0.892} & \textbf{0.820} & 0.470 & 0.805 & 0.695 & 0.449 & 0.741 & 0.683 \\ \midrule

% problem-solution nuggets row
\multirow{5}{*}{\shortstack[l]{Stack \\Overflow \\Nuggets}}
 & BM25 & 0.467 & 0.739 & 0.445 & 0.519 & 0.796 & 0.657 & 0.540 & 0.840 & 0.654 & 0.485 & 0.787 & 0.536 & 0.428 & 0.680 & 0.489 \\
 & BGE (Gemma-2) & 0.308 & 0.667 & 0.405 & 0.461 & 0.784 & 0.572 & 0.448 & 0.806 & 0.666 & 0.393 & 0.756 & 0.536 & 0.335 & 0.664 & 0.555 \\
 & E5 Mistral (7B) & 0.323 & 0.684 & 0.432 & 0.437 & 0.737 & 0.554 & 0.287 & 0.631 & 0.533 & 0.346 & 0.670 & 0.470 & 0.292 & 0.596 & 0.494 \\
 & Voyage-large-2 & 0.419 & 0.763 & 0.508 & 0.430 & 0.845 & 0.675 & 0.409 & 0.791 & 0.624 & 0.406 & 0.733 & 0.533 & 0.353 & 0.715 & 0.590 \\
 & Fusion (4 models) & \textbf{0.519} & \textbf{0.881} & \textbf{0.655} & \textbf{0.601} & \textbf{0.876} & \textbf{0.825} & \textbf{0.566} & 0.888 & 0.818 & \textbf{0.544} & \textbf{0.881} & \textbf{0.756} & \textbf{0.476} & \textbf{0.814} & \textbf{0.719} \\ \bottomrule
 
\end{tabular}
}
\caption{Pooling results by retrieval baselines (including fusion) in inference or oracle settings during \freshstack dataset construction. $\alpha$-N@10 denotes $\alpha$-nDCG@10, C@20 denotes Coverage@20 and R@50 denotes Recall@50. Stack Overflow Answer \& Nuggets both rely on the gold answer for retrieval (oracle setting), whereas other methods do not rely on the gold answer for retrieval (inference setting). Overall, we highlight the best result in \textbf{bold} for each setting.}
\label{tab:internal-results}
\end{table*}

\subsection{Retrieval and RAG Evaluation Metrics}\label{sec:eval-metrics}
IR evaluation traditionally follows the Cranfield paradigm \cite{vorhees:2009}, focusing on individual document relevance, independent of other documents. 
This is used to construct standard test collections, such as BEIR \cite{thakur:2021} and TREC datasets such as the Deep Learning (DL) track \cite{craswell:2022,craswell:2023,craswell:2024}. 
However, diversity in search~\cite{carbonell:1998, santos:2015, wu:2024} penalizes information redundancy within retrieved documents to enrich information content and improve efficiency. 
Therefore, we evaluate retrieval systems with three metrics with relevance judgments at the nugget-level: $\alpha$-nDCG@10 for diversity and relevance, Coverage@20 for nugget coverage, and Recall@50 for traditional relevancy.
For RAG evaluation, we compute the All Strict ($A_{strict}$) metric for nugget-based recall taken from TREC RAG 2024~\cite{pradeep:2024a, pradeep2025greatnuggetrecallautomating}, calculating how many nuggets are supported within a system’s response, where each nugget highlights a different aspect of the answer. Please refer to \autoref{retrieval-evaluation-metrics} for a detailed overview of each metric. 

\section{Pooling \& Qualitative Evaluation}\label{sec:internal-freshstack-eval}
In this section, we attempt to answer RQ2 by evaluating methods that retrieve documents contributing to the judgment pools. 
We first evaluate the retrieval baselines during nugget-level support judgment (or sampling pools) with both inference and oracle settings. 

In \freshstack, we are constructing a \emph{test evaluation dataset}. Therefore, we can use the Stack Overflow answer or its variants in constructing judgment pools, as discussed previously in Section \ref{sec:retrieval}. 
We pool and sample documents from different systems and techniques, similar to how existing question answering datasets are constructed, such as Natural Questions \cite{kwiatkowski:2019} or XOR-TyDI \cite{asai:2021}, which assessed the document-level relevance by calculating the answer overlap in the document.

\textbf{Experimental Settings.}~We perform retrieval with four techniques and baselines (as explained in Section \ref{sec:retrieval}) and an ensemble fusion of the top 100 documents, with each model score normalized and summed up. 
Evaluation metrics include $\alpha$-nDCG@10, Coverage@20, and Recall@50.
We use GPT-4o with a temperature setting of 0.1\footnote{Separately, we tested temperatures of 0.1 and 0.7, observing an identical downstream retrieval accuracy during \freshstack construction.} for both the automatic stages. 
Nugget generation uses a grading notes prompt with the question and answer, and support assessment uses a chain-of-thought prompt~\cite{wei:2023}, judging up to a maximum of 20 documents simultaneously with a list of nuggets generated for each question. 
Finally, we sample and judge the top 20 fusion documents from each technique and setting (including the question) to avoid sampling holes, highlighting the importance of document diversity in our judgment pools.

\subsection{Document Judgment Pooling Results}
We outline the results achieved on document judgment pools during \freshstack construction with techniques from both inference and oracle settings. Key takeaways and findings are discussed below:

\textbf{Overall Highlights.} \autoref{tab:internal-results} reveals two key findings:\ (1) Techniques in the oracle setting significantly outperform techniques from the inference setting. We observe that both the Stack Overflow answer and nuggets techniques help pool documents relevant to the question, and (2) fusion outperforms all individual models, highlighting the value of diversity in model choice, aiding in the construction of our judgment pools in niche domains and topics. 

Within the inference setting in \autoref{tab:internal-results}, we observe GPT-4o Sub-Questions achieves the best pooling results for four topics (except Yolo v7~\&~v8), showing that decomposing the question into smaller sub-questions is useful in retrieving relevant documents. 
Stack Overflow Nuggets achieve the best results (except Laravel 10~\&~11) in the oracle setting, showing that breaking down the answer into facts or nuggets is crucial. 
Amongst the individual models, BM25 achieves the best $\alpha$-nDCG@10 on all topics, asserting the importance of lexical approaches in judgment pool construction.

\subsection{Qualitative Analysis}\label{sec:quality-eval}
In our work, a crucial component is the automatic construction of nuggets and nugget-level document judgments with GPT-4o. 
To assess GPT-4o's accuracy, we calibrate with an expert human assessor (ML researcher) on a subset of LangChain, evaluating the quality of generated nuggets and nugget-document support labels for 60 randomly sampled questions.

\begin{wraptable}{r}{0.45\textwidth}
\centering
\resizebox{0.45\textwidth}{!}{
\begin{tabular}{lr lr}
\toprule
\multicolumn{2}{c}{\textbf{Nugget Quality}} & \multicolumn{2}{c}{\textbf{Judgment Quality}} \\ \midrule
Precision & 90.1 \% & Relevant & 71.7 \%\\
Recall & 96.6 \% & Partially Relevant & 11.7 \%\\
Groundedness & 96.4 \% &  Non-Relevant & 16.6 \%\\ 
\bottomrule
\end{tabular}
}
\caption{Expert evaluation of GPT-4o nugget quality and nugget-document relevance judgments on 60 sampled queries in LangChain.}
\label{tab:quality-checks}
\end{wraptable}

\subsubsection{Nugget Quality Evaluation}
For nugget quality evaluation, we ask the human assessor to answer the following questions ($A$, $B$, and $C$) after reading the Stack Overflow question, answer, and list of nuggets. 
(1) $A$: Does the nugget produce hallucinated content? requiring a boolean response 
(2) $B$: Is the information provided in the nugget minor or redundant? also requiring a boolean response.
After finishing $A$ and $B$, we ask (3) $C$: How many additional nuggets are required to cover all key ideas, requiring an integer in the response.

\textbf{Evaluation Metrics.} We measure the nugget quality by calculating three metrics, by evaluating: (1) precision (P): whether nuggets generated are accurate, (2) recall (R): whether nuggets cover the key aspects of the answer and (3) groundedness (G): whether nuggets produce non-hallucinated content, i.e., within the scope of the answer. More formally, we define them as follows:
\begin{eqnarray}
\textbf{P} = \dfrac{|Nuggets| - \mathrm{sum}(B)}{|Nuggets|}, 
\textbf{R} = \dfrac{|Nuggets| - \mathrm{sum}(B)}{|Nuggets| - \mathrm{sum}(B) + C}, 
\textbf{G} = \dfrac{|Nuggets| - \mathrm{sum}(A)}{|Nuggets|}
\end{eqnarray}
where $|Nuggets|$ denotes the count of nuggets for a given question.

\textbf{Experimental Results.} As shown in \autoref{tab:quality-checks}, nuggets in the topic of LangChain achieve above 90\% in precision and 96\% in recall and groundedness, indicating GPT-4o can generate high-quality nuggets required in the \freshstack framework. 
Most nuggets are well-grounded, i.e., do not produce hallucinated content (3.6\% error), and cover the key aspects of the answer in terms of recall (3.4\% error). 
Precision errors are higher (9.9\% error), showing nuggets may contain either minor or repeated information. Within these errors, the last positioned nugget is not informative in almost 50\% of all error cases, and either the first or second positioned nugget in the rest of the error cases.

\subsubsection{Relevance Judgment Quality Evaluation} 
We assess the relevance between nuggets and documents in nugget-level support. Since judging all documents (including negatives) for each nugget is cumbersome, we qualitatively check for precision by evaluating only the relevant pairs. 
We sample one positive document per question, totaling 60 randomly sampled nugget-document pairs. 
The annotator labels the relevance on a three-level scale: relevant, partially relevant, or non-relevant.

\textbf{Experimental Results.} As shown in \autoref{tab:quality-checks}, 71.7\% of the judged nuggets and documents are relevant, including an additional 11.7\% which are labeled partially relevant, indicating a high precision in GPT-4o support judgment. 
On the other hand, GPT-4o makes a mistake in judgment for 16.6\% of the total questions. 
This discrepancy arises from several factors: some documents are relevant to only part of the nugget's information, leading to mislabeling; ambiguity within the nugget content can cause misjudgments; and occasionally, literal grounding of a document in the nugget does not translate to semantic relevance in answering the question.

\begin{table*}[t!]

\resizebox{\textwidth}{!}{
\begin{tabular}{l|ccc|ccc|ccc|ccc|ccc}
\toprule
\multicolumn{1}{c}{\multirow{2}{*}{\textbf{Model}}} & \multicolumn{3}{c}{\textbf{LangChain}} & \multicolumn{3}{c}{\textbf{Yolo v7 \& v8}} & \multicolumn{3}{c}{\textbf{Laravel 10 \& 11}} & \multicolumn{3}{c}{\textbf{Angular 16, 17 \& 18}} & \multicolumn{3}{c}{\textbf{Godot4}} \\ 
\cmidrule(lr){2-4} \cmidrule(lr){5-7} \cmidrule(lr){8-10} \cmidrule(lr){11-13} \cmidrule(lr){14-16}
\multicolumn{1}{l}{} & $\alpha$N@10 & C@20 & \multicolumn{1}{c}{R@50} & $\alpha$N@10 & C@20 & \multicolumn{1}{c}{R@50} & $\alpha$N@10 & C@20 & \multicolumn{1}{c}{R@50} & $\alpha$N@10 & C@20 & \multicolumn{1}{c}{R@50} & $\alpha$N@10 & C@20 & R@50 \\ \midrule
 \rowcolor{paleaqua} \multicolumn{16}{l}{\textit{Inference Setting: Retrieving documents using only the Stack Overflow (SO) query.}} \\ \midrule
 BM25 & 0.230 & 0.475 & 0.261 & 0.137 & 0.342 & 0.337 & 0.319 & 0.602 & 0.441 & 0.259 & 0.551 & 0.340 & 0.144 & 0.268 & 0.200 \\
 BM25 + Reranker & \hlgreen{0.322} & \hlgreen{0.587} & \hlgreen{0.294} & \hlgreen{0.337} & \hlgreen{0.590} & \hlgreen{0.424} & \hlgreen{0.414} & \hlgreen{0.729} & \hlgreen{0.509} & \hlgreen{0.346} & \hlgreen{0.647} & \hlgreen{0.385} & \hlgreen{0.251} & \hlgreen{0.407} & \hlgreen{0.244} \\ \midrule
 BGE (Gemma-2) & 0.216 & 0.548 & 0.337 & 0.258 & 0.547 & 0.430 & 0.348 & 0.699 & 0.574 & 0.323 & 0.571 & 0.378 & 0.199 & 0.479 & 0.419 \\
  BGE (Gemma-2) + Reranker & \hlgreen{0.349} & \hlgreen{0.662} & \hlgreen{0.387} & \hlgreen{0.388} & \hlgreen{0.666} & \hlgreen{0.459} & \hlred{0.306} & \hlred{0.646} & \hlred{0.571} & \hlred{0.296} & \hlgreen{0.595} & \hlgreen{0.387} & \hlgreen{0.324} & \hlgreen{0.576} & \hlgreen{0.471} \\ \midrule
 E5 Mistral (7B) & 0.304 & 0.654 & 0.393 & 0.243 & 0.552 & 0.394 & 0.250 & 0.565 & 0.470 & 0.262 & 0.548 & 0.368 & 0.217 & 0.444 & 0.359 \\
 E5 Mistral (7B) + Reranker & \hlgreen{0.385} & \hlgreen{0.701} & \hlgreen{0.439} & \hlgreen{0.364} & \hlgreen{0.628} & \hlgreen{0.468} & \hlgreen{0.305} & \hlgreen{0.613} & \hlgreen{0.510} & \hlgreen{0.306} & \hlgreen{0.601} & \hlgreen{0.375} & \hlgreen{0.315} & \hlgreen{0.566} & \hlgreen{0.426} \\ \midrule
 Voyage-large-2 & 0.246 & 0.528 & 0.309 & 0.270 & 0.570 & 0.453 & 0.345 & 0.701 & 0.543 & 0.304 & 0.625 & 0.427 & 0.282 & 0.522 & 0.458 \\
 Voyage-large-2 + Reranker & \hlgreen{0.345} & \hlgreen{0.648} & \hlgreen{0.355} & \hlgreen{\textbf{0.418}} & \hlgreen{0.670} & \hlgreen{0.514} & \hlred{0.302} & \hlred{0.653} & \hlred{0.529} & \hlred{0.300} & \hlred{0.600} & \hlred{0.414} & \hlgreen{\textbf{0.342}} & \hlgreen{0.598} & \hlgreen{0.511} \\ \midrule
 Fusion (4 models) & 0.337 & 0.700 & 0.477 & 0.304 & 0.627 & 0.534 & \textbf{0.426} & \textbf{0.748} & \textbf{0.646} & \textbf{0.385} & \textbf{0.719} & \textbf{0.532} & 0.265 & 0.550 & 0.505 \\ 
 Fusion (4 models) + Reranker & \hlgreen{\textbf{0.397}} & \hlgreen{\textbf{0.729}} & \hlgreen{\textbf{0.501}} & \hlgreen{0.416} & \hlgreen{\textbf{0.733}} & \hlgreen{\textbf{0.592}} & \hlred{0.319} & \hlred{0.671} & \hlred{0.614} & \hlred{0.318} & \hlred{0.641} & \hlred{0.488} & \hlgreen{0.340} & \hlgreen{\textbf{0.627}} & \hlgreen{\textbf{0.545}} \\ \midrule
 \rowcolor{paleaqua} \multicolumn{16}{l}{\textit{Best Scores in the Oracle Setting taken from \autoref{tab:internal-results}: Upper Baselines on the \freshstack dataset}} \\ \midrule
SO Answer: Fusion (4 models) & 0.484 & 0.821 & 0.619 & 0.546 & 0.854 & 0.788 & 0.564 & 0.892 & 0.820 & 0.470 & 0.805 & 0.695 & 0.449 & 0.741 & 0.683 \\
SO Nuggets: Fusion (4 models) & 0.519 & 0.881 & 0.655 & 0.601 & 0.876 & 0.825 & 0.566 & 0.888 & 0.818 & 0.544 & 0.881 & 0.756 & 0.476 & 0.814 & 0.719 \\

 \midrule
 
\end{tabular}
}
\caption{Document retrieval results on \freshstack with retrieval and reranker baselines (including fusion). Best scores or upper baselines in the oracle setting are taken from \autoref{tab:internal-results}. The reranker is the Voyage AI rerank-2 model~\cite{voyage_rerank_2} reranking the top 100 documents. If the reranker improves upon the retrieval model, it is highlighted in \hlgreen{green} else \hlred{red}. We highlight the best result in \textbf{bold}.}
\label{tab:main-results}
\end{table*}

\vspace{-2mm}
\section{Main Experiments}\label{sec:external-results}
In this section, we evaluate retrievers and rerankers on document retrieval and RAG settings on the constructed \freshstack datasets, addressing RQ3 posed in our introduction. 
All models are evaluated using only the Stack Overflow question to retrieve documents in the inference setting, and do not include any information about the Stack Overflow answer or nuggets, ensuring a fair assessment. 

\textbf{Experimental Settings.} We evaluate the same retrieval models used as baselines during pooling in \freshstack: BM25, BGE (Gemma-2), E5 Mistral 7B, Voyage-large-2, and Fusion. 
In addition, we evaluate the Voyage AI rerank-2~\cite{voyage_rerank_2} as the reranker with a 16K context length, reranking the top 100 documents retrieved from each first-stage retrieval system and fusion. 
Metrics used for evaluation are defined in Section \ref{sec:eval-metrics}: $\alpha$-nDCG@10, Coverage@20, and Recall@50. 

For RAG evaluation, we generate a RAG answer naively with five LLM generators: GPT-4o-mini, GPT-4o, GPT-4.1 (nano, mini), and GPT-4.1. We feed the query and the top 20 retrieved documents concatenated together as context. Next, we evaluate whether the RAG answer supports each nugget generated in Section \ref{sec:nuggetization}, following \citet{pradeep2025greatnuggetrecallautomating}, providing three labels: \texttt{support}, \texttt{partial\_support}, or \texttt{no\_support}. We compute the All Strict ($A_{strict}$) metric for RAG evaluation.

\subsection{Document Retrieval Results}
% \autoref{tab:main-results} shows the results for each retrieval baseline without and with VoyageAI rerank-2 reranking queries on each topic. 
% Each model category is color-coded. 
% We plot the $\alpha$-nDCG@10, Coverage@20, and Recall@50 scores from left to right. 
% Key takeaways and findings are discussed below:

\textbf{Accuracy gap between oracle indicates plenty of headroom.} From \autoref{tab:main-results}, we observe techniques from the oracle setting (using Stack Overflow answers or nuggets) achieve a substantially higher $\alpha$-nDCG@10, Coverage@20, and Recall@50 in contrast to all models, including ensemble fusion and reranking with VoyageAI rerank-2. This highlights the complexity of answering \freshstack questions and demonstrates the headroom for improvement in existing code-mixed retrieval models to decrease the gap between retrieval models at inference and oracle approaches.

\textbf{Ensemble fusion outperforms individual models.} Individual retrieval models demonstrate limited success on the \freshstack dataset; whereas, the ensemble fusion of four retrieval models outperforms each retrieval model across all metrics ($\alpha$-nDCG@10, Coverage@20, and Recall@50) and all five topics, except $\alpha$-nDCG@10 on Godot4.
This highlights a crucial point: a compound retrieval system~\cite{compound-ai-blog}, developed as an ensemble of retrieval models or something similar, is required to retrieve documents for niche and challenging topics, at present. 
However, fusion is inefficient at inference time, as it adds up individual model inferences, requiring alternatives.

\textbf{Opportunities to improve reranking.}
When using a weak first-stage retrieval, neural rerankers typically improve document ranking~\cite{thakur:2021}, although it has been recently shown that this is not always the case when a strong first-stage retrieval is used~\cite{bendersky:2022,jacob:2024}. Consistent with these recent observations, reranking provides benefits over BM25 for all topics in the \freshstack dataset. 
However, for our dense retrievers, reranking provides a clear benefit on some but not all datasets. 
Specifically, while the reranker enhances $\alpha$-nDCG@10, Coverage@20, and Recall@50 for LangChain, Yolo v7~\&~v8, and Godot4, it reduces those metrics on Laravel 10~\&~11 and Angular 16,~17~\&~18 for BGE (Gemma-2), Voyage-large-2 and fusion. 
We suspect the reranker is better in certain programming languages such as Python, and we keep it as future work to understand the limitations of the neural reranker~\cite{jacob:2024}. 

\begin{table*}[t!]

\resizebox{\textwidth}{!}{
\begin{tabular}{lcc|ccc|ccc|ccc|ccc|ccc}
\toprule
\multicolumn{1}{c}{\multirow{3}{*}{\textbf{Technique}}} & \multicolumn{1}{c}{\multirow{3}{*}{\textbf{Retrieval}}} & \multicolumn{1}{c}{\multirow{3}{*}{\textbf{Generator}}} & \multicolumn{3}{c}{\textbf{LangChain}} & \multicolumn{3}{c}{\textbf{Yolo v7 \& v8}} & \multicolumn{3}{c}{\textbf{Laravel 10 \& 11}} & \multicolumn{3}{c}{\textbf{Angular 16, 17 \& 18}} & \multicolumn{3}{c}{\textbf{Godot4}} \\ 
\cmidrule(lr){4-6} \cmidrule(lr){7-9} \cmidrule(lr){10-12} \cmidrule(lr){13-15} \cmidrule(lr){16-18}
\multicolumn{1}{l}{} & \multicolumn{1}{l}{} & \multicolumn{1}{l}{} & nano & mini & full & nano & mini & full & nano & mini & full & nano & mini & full & nano & mini & full \\ \midrule
\rowcolor{paleaqua} \multicolumn{18}{l}{\textit{Inference Setting: Retrieving documents using only the Stack Overflow query.}} \\ \midrule
Closed & No Retrieval & GPT-4o & -- & 0.395 &  0.524 & -- & 0.461 & 0.591 & -- & 0.512 & 0.574 & -- & 0.486 & 0.568 & -- &   0.415 & 0.518 \\
Book & No Retrieval & GPT-4.1 & 0.444 & 0.517 & 0.564 & 0.470 & 0.663 & 0.647 & 0.557 & 0.646 & 0.621 & 0.508 & 0.604 & 0.597 & 0.483 & 0.616 & 0.573 \\ \midrule
& Fusion & GPT-4o & -- &  0.464 & 0.568 & -- & 0.477 & 0.630 & -- & 0.557 & 0.635 & -- & 
 0.536 & 0.629 & -- &  0.452 & 0.544 \\
StackOverflow & Fusion & GPT-4.1 & 0.438 & 0.578 & 0.610 & 0.571 & 0.649 & 0.624 & 0.572 & 0.668 & 0.660 & 0.575 & 0.670 & 0.674 & \textbf{0.492} & 0.573 & 0.595 \\
Query & Fusion + Rerank & GPT-4o & -- & 0.444 & 0.587 & -- & 0.492 & 0.625 & -- & 0.545 & 0.617 & -- & 0.551 & 0.620 & -- & 0.428 & 0.551 \\
 & Fusion + Rerank & GPT-4.1 & 0.467 & 0.594 & 0.625 & 0.527 & \textbf{0.679} & \textbf{0.684} & 0.583 & 0.657 & 0.651 & 0.564 & 0.663 & 0.650 & 0.491 & 0.578 & 0.591 \\
\midrule
\rowcolor{paleaqua} \multicolumn{18}{l}{\textit{Oracle Setting (Upper Baseline): Using the Stack Overflow answer directly or its variants for retrieval of documents within the corpus}} \\ \midrule
StackOverflow & Fusion & GPT-4o & -- & 0.492 & 0.618 & -- & 0.559 & 0.668 & -- & 0.549 & 0.656 & --  &  0.584 &  0.680 & -- & 0.477 & 0.576 \\
Nuggets & Fusion & GPT-4.1 & \textbf{0.533} & \textbf{0.654} & \textbf{0.651} & \textbf{0.591} & 0.667 & 0.667 & \textbf{0.607} & \textbf{0.681} & \textbf{0.696} & \textbf{0.626} & \textbf{0.717} & \textbf{0.709} & 0.489 & \textbf{0.628} & \textbf{0.668} \\ \bottomrule
\end{tabular}
}
\caption{RAG evaluation results measuring nugget recall with All Strict ($A_{strict}$) scores on LLM-generated answer with: GPT-4o-mini and GPT-4o, GPT-4.1 (nano, mini) and GPT-4.1. The knowledge cutoff date for GPT-4o series is October 2023 and GPT-4.1 series is June 2024.}
\label{tab:rag-results}
\end{table*}

\subsection{RAG Evaluation Results} 

\paragraph{Retrieved context is key for RAG accuracy.} 
From the evaluation results measuring $A_{strict}$ capturing nugget recall in \autoref{tab:rag-results}, the RAG answer generated in the oracle setting outperforms other techniques except Yolo v7~\&~v8, indicating that the oracle retrieved context is the key. Consistent across all observations, GPT-4.1 performs better than GPT-4o across all versions and topics, which we suspect is likely due to a more recent knowledge cutoff date (June 2024 versus October 2023). We likely suspect this as a possibility for GPT-4.1 (especially, mini) surprisingly high effectiveness of the closed-book answer.
Lastly, the fusion + rerank setting baseline is competitive, outperforming even the oracle setting in Yolo v7~\&~v8. 

\vspace{-2mm}
\section{Conclusion} 
The emergence of RAG has improved modern retrieval systems by allowing real-time data incorporation into LLMs. 
However, existing IR and RAG benchmarks that measure retrieval quality are outdated. 
In this work, we introduce a holistic framework, \freshstack, to construct challenging datasets to evaluate retrieval systems realistically. We source real user questions and answers from Stack Overflow and build a document corpus using technical documents from public GitHub repositories. 
Using \freshstack, we construct datasets on five niche topics and evaluate four frontier retrieval models and a reranker model in the document retrieval setting. 
The accuracy gap observed between the retrieval models and approaches in the oracle setting indicates plenty of headroom for improvement, and we identify cases that may motivate future research in reranking. 
We hope \freshstack will encourage the community to build more challenging and realistic IR and RAG datasets in the future.

% \begin{ack}
% We thank Sean Kulinski and Alexander Trott for helping us set up the grading notes prompt required in nugget generation. We also thank Jacob Portes, Max Marion, Matei Zaharia, and others from Databricks who provided feedback at the early stages of the project.
% \end{ack}

\clearpage
\bibliography{neurips_2025}
\bibliographystyle{acl_natbib}
\clearpage

%%%%%%%%%%%%%%%%%%%%%%%%%%%%%%%%%%%%%%%%%%%%%%%%%%%%%%%%%%%%

\appendix

\section{Technical Appendices and Supplementary Material}

\begin{wraptable}{r}{0.5\textwidth}
\centering
\resizebox{0.5\textwidth}{!}{
\begin{tabular}{lcccc}
\toprule
\multirow{2}{*}{\textbf{IR~/~RAG Benchmarks}} & \textbf{Niche} & \textbf{Complex} & \textbf{Dynamic} & \textbf{Challenge} \\
 & \textbf{Domains} & \textbf{Questions} & \textbf{Updates} & \textbf{Level} \\
\midrule
CQADupstack~\cite{hoogeveen:2015} & No & No & No & Easy \\
CodeSearchNet~\cite{husain:2019} & No & No & No & Easy \\
COIR~\cite{li:2024} & Limited & Yes & No & Moderate \\
Stack Overflow-QA~\cite{li:2024} & No & Yes & No & Moderate \\
CodeRAG-Bench~\cite{wang:2024} & Limited & No & No & Moderate \\
Neural Code Search~\cite{li:2019b} & No & No & No & Moderate \\
SWE-Bench~\cite{jimenez:2024} & No & Yes & Yes & High \\
\rowcolor{paleaqua} \textbf{\freshstack (ours)} & \textbf{Yes} & \textbf{Yes} & \textbf{Yes} & \textbf{High} \\ 
\bottomrule
\end{tabular}}
\caption{A comparison of existing IR/RAG evaluation benchmarks with \freshstack.}
\label{tab:dataset_comparison}
\end{wraptable}
\section{Comparison of \freshstack with Existing IR and RAG Benchmarks}

\autoref{tab:dataset_comparison} compares \freshstack against existing code-focused IR or RAG benchmarks. Below, we briefly describe a few advantages of \freshstack over existing RAG benchmarks:

First, the \freshstack framework utilizes user-asked questions and curated answers, making the evaluation challenging. A majority of existing benchmarks are unrealistic, derived from easily retrievable topics and queries, such as Neural Code Search~\cite{li:2019b}, making it easy to retrieve and answer them, rather than being grounded in solving real user problems provided in \freshstack.
We are not crafting artificial (or LLM-generated) questions or sampling questions myopically. 
Second, all answers in \freshstack are supported in real time by information from technical documentation in GitHub repositories. 
Third, the framework is designed to be general and scalable without modification. 
Finally, \freshstack is focused on niche domains and recent topics, taking careful measures to mitigate risks with data contamination introduced by LLMs, ensuring that the benchmark is not susceptible to distortion or leaderboard overfitting~\cite{singh2025leaderboardillusion}.

\section{Retrieval and RAG Evaluation Metrics}\label{retrieval-evaluation-metrics}

\subsection{Retrieval Evaluation Metrics}
\textbf{$\alpha$-nDCG@k.} Introduced by \citet{clarke:2008}, this variant of Normalized Discounted Cumulative Gain (nDCG) measures search diversification. 
The $\alpha$ parameter is a geometric penalization for redundant documents, i.e., each redundant document achieves a penalization of $\times(1-\alpha)$. 
Despite the metric being used for different user intents, we utilize it to ensure document rankings reference diverse nuggets in the answer. We would ask the reader to refer to ~\citet{clarke:2008} for more information.

\textbf{Coverage@k.} The metric introduced in our work measures the average proportion of the nuggets covered by the top-k retrieved documents. The mathematical formula is calculated as:
\begin{equation}
\text{Coverage@k} = \frac{1}{|Q|} \sum_{q=1}^{Q} \frac{\left| \bigcup_{i=1}^{k} \text{Nuggets}(d_{qi}) \right|}{\left| \text{Nuggets}(q) \right|}
\end{equation}
where $Q$ contains all questions, $\text{Nuggets}(d_{qi})$ are nuggets supported by document $d_{qi}$ and $\text{Nuggets}(q)$ are nuggets for question $q$.

\textbf{Recall@k.} The standard relevance metric measures the proportion of relevant documents retrieved within the top-k results, out of all relevant documents for a given question. A document is judged relevant if it supports at least one nugget.

\subsection{RAG Evaluation Metric}
\paragraph{All Strict ($A_{strict}$)} 
Introduced in \citet{pradeep2025greatnuggetrecallautomating}, for each query, we have a list of nuggets generated from Section \ref{sec:nuggetization}, and for each RAG answer generated, we have a record of which nuggets it contains, in terms of a three-way judgment:\ \texttt{support}, \texttt{partial\_support}, and \texttt{no\_support}. The final step is to compute the score for the RAG answer to a query $q$.
The score of a run is simply the mean of the score for each query. We compute the following scores per query:

We calculate a score based on all nuggets in the RAG answer, but with strict nugget matching. For a given nugget $i$:
\begin{align}
    p_i &= \begin{cases}
        1 & \text{if assignment = \texttt{support}} \\
        0 & \text{otherwise}
    \end{cases}
\end{align}
The ``All Strict'' score is then calculated as:
$$A_{strict} = \frac{\sum_i p_i}{|Nuggets|},$$ 
where $|Nuggets|$ denotes the count of nuggets for a given query $q$.

\section{\freshstack Instance Description}
Each \freshstack dataset instance contains the following four components, as shown in \autoref{fig:langchain}. A complete example of a dataset instance is shown in \autoref{tab:example}.
\begin{itemize}[leftmargin=*]
    \item \textbf{Question \& Answer}: The title and body (description) of the Stack Overflow post as the question, with the accepted answer. The title is a short sentence, and the body contains the detailed issue with code snippets and/or outputs.
    \item \textbf{Nuggets}: The list of atomic facts highlighting the essential information in the Stack Overflow question and answer.
    \item \textbf{Document Corpus}: The exhaustive list of chunked source documents (code snippets, text documentation, etc.) compiled from GitHub repositories.
    \item \textbf{Relevance Judgments}: Unlike traditional IR benchmarks, such as BEIR \cite{thakur:2021}, which contain question and document-level relevance judgments, \freshstack datasets contain nugget-level relevance judgments for document chunks.
\end{itemize}

\section{Discussion and Future Work}\label{sec:discussion}

\freshstack is a holistic framework for building challenging IR and RAG evaluation datasets. We apply the framework to community-sourced questions (with curated answers) and documents sourced from GitHub repositories. The framework is adaptable to other domains like Stack Exchange or internal forums. 

While we benchmarked a few key retrieval models in our work, we would like to benchmark code-focused retrieval models like voyage-3~\cite{voyage_3}, Code-T5+~\cite{wang:2023}, CodeRankEmbed~\cite{suresh:2024}, and Jina-Code-v2~\cite{gunther:2023}, and rerankers such as CodeRankLLM~\cite{suresh:2024}, in the future.

\textbf{Answer Evaluation.} We focused on the evaluation of the retrieval setting primarily due to two reasons: (1) existing RAG datasets evaluate retrieval using relevance criteria only, however, we evaluated models based on both diversity and relevance criteria, and (2) a crucial step in \freshstack is sourcing and building a document corpus and developing a general framework for high-quality pools and automatic judgments, which we can evaluate better in the retrieval setting. We evaluated the quality of LLM-based answer generation with nugget-based recall with $A_{strict}$ metric~\cite{pradeep:2024a, pradeep2025greatnuggetrecallautomating}, which calculates how many nuggets are supported within a system's response. However, we keep an in-depth evaluation of the RAG answer, accounting for other factors such as fluency or support~\cite{thakur2025supportevaluationtrec2024}, as future work.

\begin{figure*}[h]
    \centering
    % langchain
    \includegraphics[width=0.6\textwidth,clip, trim=0 0 0 0]{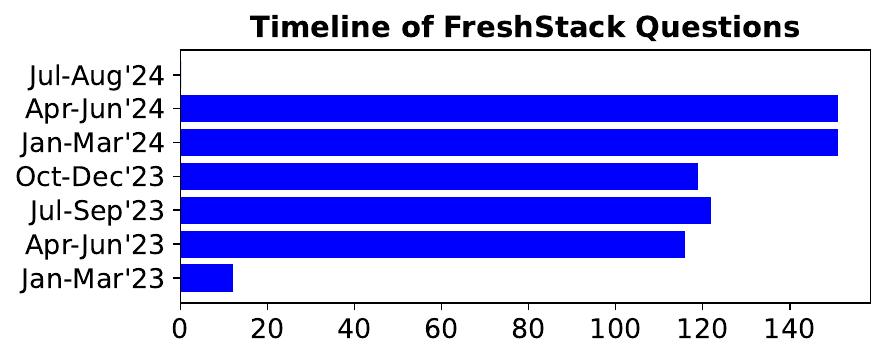}
  \caption{Timeline versus frequency of how many \freshstack queries were asked on Stack Overflow in every quarter. All queries included in \freshstack were asked between January 2023 and June 2024, with the highest frequencies observed in 2024, showing the growing importance of all five topics.}
  \label{fig:freshstack-questions}
\end{figure*}

\textbf{Benchmark Contamination.} The \freshstack dataset is built on Stack Overflow data, making it susceptible to future data contamination. Newer released LLMs such as GPT-4.1 with a recent knowledge cutoff date\footnote{\url{https://help.openai.com/en/articles/9624314-model-release-notes}} of June 2024, provide the possibility of dataset contamination with GPT-4.1 as \freshstack queries originated from January 2023 until June 2024 (as shown in \autoref{fig:freshstack-questions}).
To mitigate data contamination, the \freshstack framework can add newer questions in existing topics, retire old and contaminated topics, and add newer topics that develop in the future. The relevance of \freshstack in the community relies on a continued commitment to keeping it updated in the upcoming years.

\begin{figure*}[t!]
    \centering
    % langchain
    \includegraphics[width=\textwidth,clip, trim=0 0 0 0]{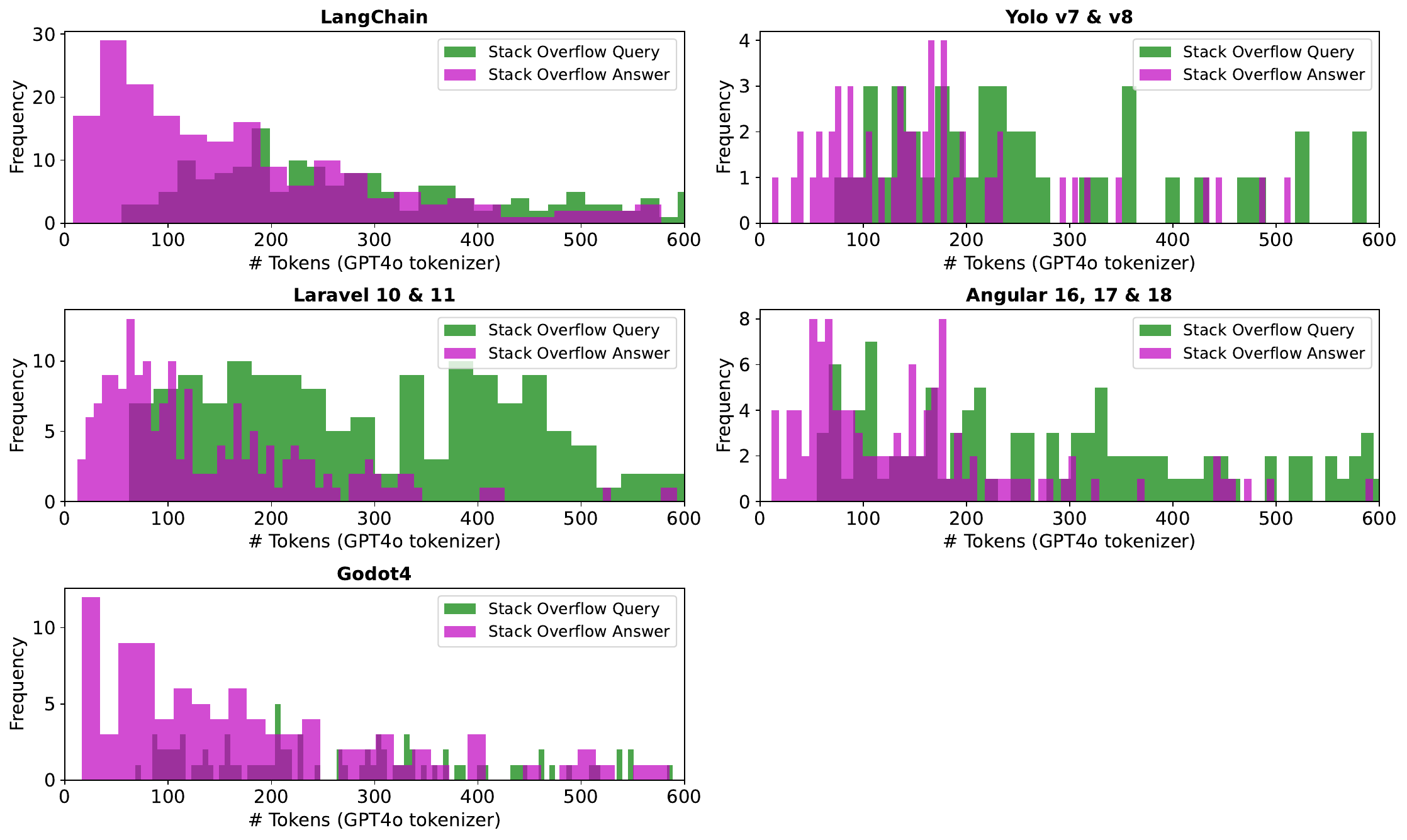}
  \caption{Token distribution of Stack Overflow questions and answers for all topics in \freshstack. Unlike other benchmarks, \freshstack questions (highlighted in green) are much longer than their answers (highlighted in maroon).}
  \label{fig:token-distribution}
\end{figure*}
\begin{table*}[h]
\centering
\footnotesize
\resizebox{\textwidth}{!}{
\begin{tabular}{@{}lp{8cm}@{}}
\toprule
\textbf{Topic} & \textbf{GitHub Repositories \& Licenses} \\
\midrule
\textbf{LangChain} &
[1] \href{https://github.com/langchain-ai/langchain}{langchain-ai/langchain}: MIT license \\ 
& 
[2] \href{https://github.com/langchain-ai/langchainjs}{langchain-ai/langchainjs}: MIT license \\ &
[3] \href{https://github.com/langchain-ai/langchain-nextjs-template}{langchain-ai/langchain-nextjs-template}: MIT license \\ &
[4] \href{https://github.com/chroma-core/chroma}{chroma-core/chroma}: Apache-2.0 license \\ &
[5] \href{https://github.com/openai/openai-cookbook}{openai/openai-cookbook}: MIT license \\ &
[6] \href{https://github.com/openai/openai-python}{openai/openai-python}: Apache-2.0 license \\ &
[7] \href{https://github.com/run-llama/llama_index}{run-llama/llama\_index}: MIT license \\ &
[8] \href{https://github.com/Azure-Samples/openai}{Azure-Samples/openai}: MIT license \\ &
[9] \href{https://github.com/Azure-Samples/azure-search-openai-demo}{Azure-Samples/azure-search-openai-demo}: MIT license \\ &
[10] \href{https://github.com/huggingface/transformers}{huggingface/transformers}: Apache-2.0 license \\ \midrule
\textbf{Yolo v7 \& v8} &
[1] \href{https://github.com/ultralytics/ultralytics}{ultralytics/ultralytics}: AGPL-3.0 license \\ &
[2] \href{https://github.com/ultralytics/docs}{ultralytics/docs}: AGPL-3.0 license \\ &
[3] \href{https://github.com/pytorch/pytorch}{pytorch/pytorch}: Modified BSD license \\ &
[4] \href{https://github.com/WongKinYiu/yolov7}{WongKinYiu/yolov7}: GPT-3.0 license \\ &
[5] \href{https://github.com/opencv/opencv}{opencv/opencv}: Apache-2.0 license \\ \midrule
\textbf{Laravel 10 \& 11} &
[1] \href{https://github.com/laravel/framework}{laravel/framework}: MIT license \\ &
[2] \href{https://github.com/laravel/laravel}{laravel/laravel}: MIT license \\ &
[3] \href{https://github.com/laravel/laravel.com}{laravel/laravel.com}: MIT license \\ &
[4] \href{https://github.com/laravel/docs}{laravel/docs}: MIT license \\ &
[5] \href{https://github.com/laravel/breeze}{laravel/breeze}: MIT license \\ &
[6] \href{https://github.com/livewire/livewire}{livewire/livewire}: MIT license \\ &
[7] \href{https://github.com/php/php-src}{php/php-src}: PHP license \\ &
[8] \href{https://github.com/php/doc-en}{php/doc-en}: PHP license \\ &
[9] \href{https://github.com/php/web-php}{php/web-php}: PHP license \\ \midrule
\textbf{Angular 16, 17 \& 18} &
[1] \href{https://github.com/angular/angular}{angular/angular}: MIT license \\ &
[2] \href{https://github.com/angular/components}{angular/components}: MIT license \\ &
[3] \href{https://github.com/angular/angular-cli}{angular/angular-cli}: MIT license \\ &
[4] \href{https://github.com/microsoft/TypeScript}{microsoft/TypeScript}: Apache-2.0 license \\ \midrule
\textbf{Godot4}  &
[1] \href{https://github.com/godotengine/godot}{godotengine/godot}: MIT license \\ &
[2] \href{https://github.com/godotengine/godot-demo-projects}{godotengine/godot-demo-projects}: MIT license \\ &
[3] \href{https://github.com/godotengine/godot-docs}{godotengine/godot-docs}: CC BY 3.0 license \\ &
[4] \href{https://github.com/godotengine/godot-website}{godotengine/godot-website}: MIT license \\ &
[5] \href{https://github.com/GDQuest/learn-gdscript}{GDQuest/learn-gdscript}: MIT license \\ &
[6] \href{https://github.com/dotnet/csharplang}{dotnet/csharplang}: GPL license \\
\bottomrule
\end{tabular}}
\caption{A list of the GitHub repositories with their licenses for constructing the document collection for each topic in \freshstack.}
\label{tab:github_repositories}
\end{table*}

\begin{table*}[h!]
\centering
\small
\resizebox{\textwidth}{!}{
\begin{tabular}{@{}p{1.5cm}p{16cm}@{}}
\toprule
\textbf{LangChain} & \textbf{Query ID: 78256389} \\
\midrule
\textbf{Stack Overflow Query} &
\textbf{Title}: Chromadb from\_documents function giving error. 

\textbf{Text:} The following function was working till a few days ago but now gives this error:
ValueError: Expected EmbeddingFunction.\_call\_ to have the following signature: odict\_keys(['self', 'input']), got odict\_keys(['args', 'kwargs']) 
Please see \url{https://docs.trychroma.com/embeddings} for details of the `EmbeddingFunction` interface. Please note the recent change to the `EmbeddingFunction` interface: \url{https://docs.trychroma.com/migration\#migration-to-0416---november-7-2023}. 
I am not sure what changes are necessary to work with this.
\begin{lstlisting}[language=Python]
def create_chromadb(link):
    embedding_function = SentenceTransformerEmbeddings(model_name="all-MiniLM-L6-v2")
    loader = TextLoader(link)
    documents = loader.load()

    # Split the documents into chunks (no changes needed here)
    text_splitter = RecursiveCharacterTextSplitter(chunk_size=1000, chunk_overlap=500)
    chunks = text_splitter.split_documents(documents)

    # Update for new EmbeddingFunction definition
    # D is set to the type of documents (Text in this case)
    D = Union[str, List[str]] # Adjust based on your document format (single string or list of strings)
    embedding_function: EmbeddingFunction[D] = embedding_function

    # Initialize Chroma with the embedding function and persist the database
    db = Chroma.from_documents(chunks, embedding_function, ids=None, collection_name="langchain", persist_directory="./chroma_db")
    db.persist()
    print(f"Saved {len(chunks)} chunks")

    return db
\end{lstlisting} \\ \midrule
\textbf{Stack Overflow Answer} &
I slightly modify your code, using `HuggingFaceEmbeddings` instead of `SentenceTransformerEmbeddings`.
\begin{lstlisting}[language=Python]
from langchain_community.embeddings import HuggingFaceEmbeddings
embedding = HuggingFaceEmbeddings(model_name="sentence-transformers/all-MiniLM-L6-v2")

from langchain_community.vectorstores import Chroma
db = Chroma.from_documents(documents=chunks, embedding=embedding, persist_directory="/tmp/chroma_db")
db.persist()
\end{lstlisting} \\ \midrule
\textbf{GPT-4o Nuggets} & 
1. The error is due to a mismatch in the function signature expected by `Chroma.from\_documents` when using `SentenceTransformerEmbeddings`.

\hlgreen{2. Use `HuggingFaceEmbeddings` instead of `SentenceTransformerEmbeddings` to resolve the error.}

3.  Import `HuggingFaceEmbeddings` from `langchain\_community.embeddings`.

\hlgreen{4. Initialize `HuggingFaceEmbeddings` with the model name ``sentence-transformers/all-MiniLM-L6-v2''.}

\hlgreen{5. Pass the initialized `HuggingFaceEmbeddings` to the `Chroma.from\_documents` function.} \\ \midrule

\textbf{Retrieved GitHub Document} & \textbf{Document ID}: langchain/templates/intel-rag-xeon/ingest.py\_0\_1486

\textbf{Supported Nuggets:} Nugget \#2, Nugget \#4, and Nugget \#5.
\begin{lstlisting}[language=Python]
import os

from langchain.text_splitter import RecursiveCharacterTextSplitter
from langchain_chroma import Chroma
from langchain_community.document_loaders import UnstructuredFileLoader
from langchain_community.embeddings import HuggingFaceEmbeddings
from langchain_core.documents import Document

def ingest_documents():
    """
    Ingest PDF to Redis from the data/ directory that
    contains Edgar 10k filings data for Nike.
    """
    # Load list of pdfs
    data_path = "data/"
    doc = [os.path.join(data_path, file) for file in os.listdir(data_path)][0]

    print("Parsing 10k filing doc for NIKE", doc)
    text_splitter = RecursiveCharacterTextSplitter(chunk_size=1500, chunk_overlap=100, add_start_index=True)
    loader = UnstructuredFileLoader(doc, mode="single", strategy="fast")
    chunks = loader.load_and_split(text_splitter)

    print("Done preprocessing. Created", len(chunks), "chunks of the original pdf")

    # Create vectorstore
    embedder = HuggingFaceEmbeddings(model_name="sentence-transformers/all-MiniLM-L6-v2")

    documents = []
    for chunk in chunks:
    doc = Document(page_content=chunk.page_content, metadata=chunk.metadata)
    documents.append(doc)

    # Add to vectorDB
    _ = Chroma.from_documents(documents=documents, collection_name="xeon-rag", embedding=embedder, persist_directory="/tmp/xeon_rag_db")


if __name__ == "__main__":
    ingest_documents()
\end{lstlisting}
\\ \bottomrule
\end{tabular}}
\caption{A complete example of a dataset instance from LangChain in \freshstack. The relevant nuggets supported by the retrieved GitHub document are highlighted in \hlgreen{green}.}
\label{tab:example}
\end{table*}

\clearpage

\end{document}